\RequirePackage{amsmath}
\documentclass{llncs}
\usepackage{amssymb,amsfonts}
\usepackage[english]{babel}
\usepackage{graphicx}
\usepackage{url}
\usepackage{multirow}

\begin{document}

\title{Faster Dual-Key Stealth Address for Blockchain-Based Internet of Things Systems}
%
%
\author{Xinxin Fan}
\authorrunning{Xinxin Fan} 
%
%
\institute{IoTeX\\
\email{xinxin@iotex.io}}

\maketitle              

\begin{abstract}
Stealth address prevents public association of a blockchain transaction's output with a recipient's wallet address and hides the actual destination address of a transaction. While stealth address provides an effective privacy-enhancing technology for a cryptocurrency network, it requires blockchain nodes to actively monitor all the transactions and compute the purported destination addresses, which restricts its application for resource-constrained environments like Internet of Things (IoT). In this paper, we propose \textsf{DKSAP-IoT}, a faster dual-key stealth address protocol for blockchain-based IoT systems. \textsf{DKSAP-IoT} utilizes a technique similar to the TLS session resumption to improve the performance and reduce the transaction size at the same time between two communication peers. Our theoretical analysis as well as the extensive experiments on an embedded computing platform demonstrate that \textsf{DKSAP-IoT} is able to reduce the computational overhead by at least 50\% when compared to the state-of-the-art scheme, thereby paving the way for its application to blockchain-based IoT systems.

\end{abstract}
\section{Introduction}
The Internet of Things (IoT) has been connecting extraordinarily large number of smart devices to the Internet and driving the digital transformation of industry. Unfortunately, existing cloud-centric IoT systems have a number of significant disadvantages such as high system maintenance costs, slow response time, security and privacy concerns, etc. Blockchain, a form of distributed, immutable and time-stamped ledger technology, has been perceived as a promising solution to address the aforementioned problems and to securely unlock the business and operational values of IoT. The combination of blockchain and IoT facilitates the sharing of services and resources, creates audit trails and enables automation of time-consuming workflows in various applications. While combining these two technologies is creating new levels of trust, the decentralized network and public verifiability of blockchain transactions often do not provide the strong security and privacy properties required by the users.

During the past few years, quite a few cryptographic techniques such as ring signature \cite{Riv:Sha:Tau}, stealth address \cite{bytecoin}, and zero-knowledge proof \cite{Gol:Mic:Rac} have been employed to ensure transaction privacy for senders, receivers and transaction amount in blockchains \cite{monero,ryn:tec,cryptonote,zcash}. This work focuses on stealth address, a privacy protection technique for receivers of cryptocurrencies. Stealth address requires the sender to create random one-time addresses for every transaction on behalf of the recipient so that different payments made to the same payee unlinkable. The most basic stealth address scheme \cite{bytecoin} was first sketched by a Bitcoin Forum member named `ByteCoin' in 2011, which was then improved in \cite{cryptonote,Todd} by introducing the random ephemeral key pair and fixing the issue that the sender might change the mind and reverse the payment. Later on, a dual-key enhancement \cite{shadow} to the previous stealth address schemes was implemented in 2014, which utilized two pairs of cryptographic keys for designated third parties (e.g., auditors, proxy servers, read-only wallets, etc.) removing the unlinkability of the stealth addresses without simultaneously allowing payments to be spent.

The dual-key stealth address protocol (\textsf{DKSAP}) provides strong anonymity for transaction receivers and enables them to receive unlinkable payments in practice. However, this approach does require blockchain nodes to constantly compute the purported destination addresses and find the corresponding matches in the blockchain. While this process works well for full-fledged computers, it poses new challenges for resource-constrained IoT devices. Considering the limited energy budget of smart devices, we propose a lightweight variant of \textsf{DKSAP}, namely \textsf{DKSAP-IoT}, which is based on the similar idea as the TLS session resumption \cite{Die08,Sal08} and requires both the sender and receiver to keep track of the continuously updated pairwise keys for each payment session. \textsf{DKSAP-IoT} is able to improve the performance of \textsf{DKSAP} by at least 50\% and reduce the transaction size simultaneously, thereby providing an efficient solution to protecting the privacy of recipients in blockchain-based IoT systems.

The rest of the paper is organized as follows: Section~\ref{sec:pre} gives a brief overview of the elliptic curve cryptography, followed by the description of the dual-address stealth address protocol (\textsf{DKSAP}) in Section~\ref{sec:dksap}. In Section~\ref{sec:dksapiot}, we present \textsf{DKSAP-IoT}, a faster dual-key stealth address protocol for blockchain-based IoT systems. Section~\ref{sec:performance} analyzes the security and performance of the proposed scheme. Finally, Section~\ref{sec:conclusion} concludes this contribution.  

\section{Preliminaries}
\label{sec:pre}

An elliptic curve $E$ over a field $\mathbb{F}$ is defined by the Weierstrass equation:
\begin{displaymath}
E(\mathbb{F}): y^2 + a_1xy + a_3y = x^3 + a_2x^2 + a_4x + a_6,
\end{displaymath}
where $a_1, a_2, a_3, a_4, a_6 \in \mathbb{F}$ and the curve discriminant $\Delta \neq 0$. The set of solutions $(x, y) \in \mathbb{F} \times \mathbb{F}$ satisfying the above equation along with the identity element $\mathcal {O}$, or point-at-infinity, form an abelian group under the addition operation $+$ (i.e., the chord-and-tangent group law). It is this abelian group that is used in the construction of elliptic curve cryptosystems. Given an elliptic curve point $G \in E(\mathbb{F})$ and an integer $k$, the scalar multiplication $kG$ is defined by the addition of the point $G$ to itself $k - 1$ times, i.e., 
\begin{displaymath} 
kG = \underbrace{G + G + \cdots + G}_\text{$k - 1$ additions}.
\end{displaymath}
The scalar multiplication is the fundamental operation in elliptic curve based cryptographic protocols such as the Elliptic Curve Diffie-Hellman (ECDH) key agreement \cite{SEC1} and the Elliptic Curve Digital Signature Algorithm (ECDSA) \cite{SEC1}, etc. The security of elliptic curve cryptosystems is based on the difficulty of solving the Elliptic Curve Discrete Logarithm Problem (ECDLP) \cite{Kob87,Mil86}. This problem involves finding the integer $k$ ($0 < k < n$) given a point $kG$, where $n$ is the group order of $E(\mathbb{F})$. The 15 elliptic curves have been recommended by NIST in the FIPS 186-2 standard for U.S. federal government \cite{FIPS}, which are also contained in the specification defined by the Standards for Efficient Cryptography Group (SECG) \cite{SEC2}. For example, the elliptic curve used in Bitcoin is called \textsf{secp256k1} with parameters specified by SECG \cite{SEC2}. For more details about elliptic curve cryptography, the interested reader is referred to \cite{Han03}.  

\section{Dual-Key Stealth Address Protocol (\textsf{DKSAP})}
\label{sec:dksap}
The first full working implementation of \textsf{DKSAP} was announced by a developer known as rynomster/sdcoin in 2014 for \textbf{ShadowSend} \cite{shadow}, a capable, efficient and decentralized anonymous wallet solution. The \textsf{DKSAP} has been realized in a number of cryptocurrency systems since then, including \textbf{Monero} \cite{monero}, \textbf{Samourai Wallet} \cite{samourai}, \textbf{TokenPay} \cite{tokenpay}, just to name a few. The protocol takes advantage of two pairs of cryptographic keys, namely a `scan key' pair and a `spend key' pair, and computes a one-time payment address per transaction, as illustrated in Fig. 1.

\begin{figure}[ht]
\centering
\label{fig:stealth}
\includegraphics[width=0.97\textwidth]{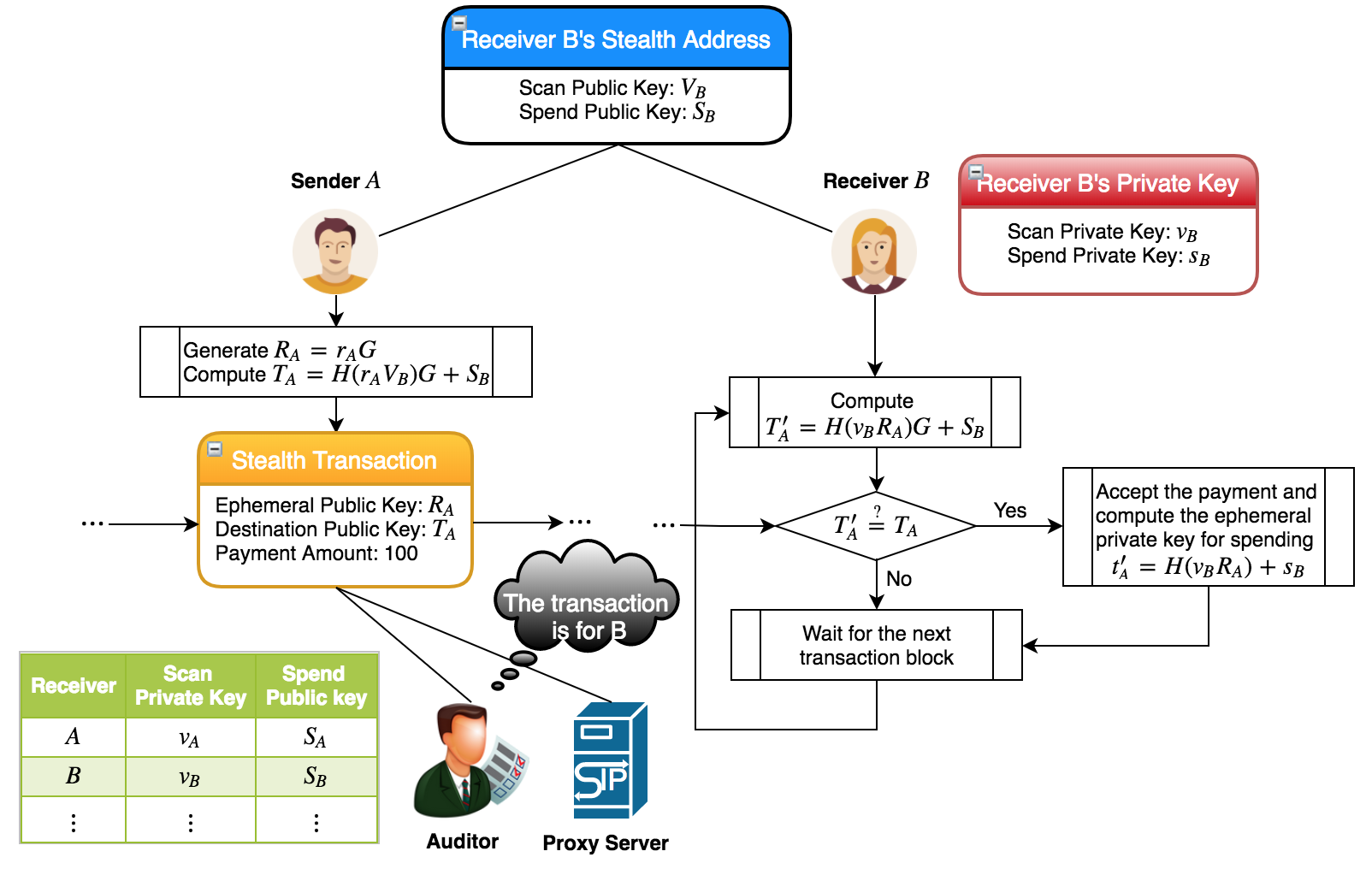}
\caption{The Dual-Key Stealth Address Protocol (\textsf{DKSAP})}
\end{figure}

When a sender $A$ would like to send a transaction to a receiver $B$ in a stealth mode \cite{shadow}, \textsf{DKSAP} works as follows:

\begin{enumerate}
\item The receiver $B$ has a pair of private/public keys $(v_B, V_B)$ and $(s_B, S_B)$, where $v_B$ and $s_B$ are called B's `scan private key' and `spend private key', respectively, whereas $V_B = v_BG$ and $S_B = s_BG$ are the corresponding public keys. Note that none of $V_B$ and $S_B$ ever appear in the blockchain and only the sender $A$ and the receiver $B$ know those keys.
\item The sender $A$ generates an ephemeral key pair $(r_A, R_A)$ with $R_A = r_AG$ and $0 < r_A < n$, and sends $R_A$ to the receiver $B$.
\item Both the sender $A$ and the receiver $B$ can perform the ECDH protocol to compute a shared secret:
\begin{displaymath}
c_{AB} = H(r_Av_BG) = H(r_AV_B) = H(v_BR_A),
\end{displaymath}
where $H(\cdot)$ is a cryptographic hash function.
\item The sender $A$ can now generate the destination address of the receiver $B$ to which $A$ should send the payment:
\begin{displaymath}
T_A = c_{AB}G + S_B.
\end{displaymath}
Note that the one-time destination address $T_A$ is publicly visible and appears on the blockchain.
\item Depending on whether the wallet is encrypted, the receiver $B$ can compute the same destination address in two different ways:
\begin{displaymath}
T_A' = c_{AB}G + S_B = (c_{AB} + s_B)G.
\end{displaymath}
The corresponding ephemeral private key is
\begin{displaymath}
t_A' = c_{AB} + s_B,
\end{displaymath}
which can only be computed by the receiver $B$, thereby enabling $B$ to spend the payment received from $A$ later on.
\end{enumerate}
In \textsf{DKSAP}, the receiver $B$ needs to actively scan the blockchain transactions, calculate the purported destination address and compare it with the one in each block until a match is found. In the case that an auditor or a proxy server exists in the system, the receiver $B$ can share the `scan private key' $v_B$ and the `spend public key' $S_B$ with the auditor/proxy server so that those entities can scan the blockchain transactions on behalf of the receiver $B$. However, they are not able to compute the ephemeral private key $t_A'$ and spend the payment received from the sender $A$.   

\section{Faster Dual-Key Stealth Address Protocol for Internet of Things ($\textsf{DKSAP-IoT}$)}
\label{sec:dksapiot}
In this section, we describe a faster dual-key stealth address protocol called $\textsf{DKSAP-IoT}$, which is dedicatedly designed for blockchain-based IoT systems.

\subsection{Design Rationale}
In $\textsf{DKSAP}$, the receiver $B$ scans the blockchain and calculates the purported destination address for each transaction, which requires computations of two scalar multiplications, including one random-point scalar multiplication with the ephemeral public key $R_A$ and one fixed-point scalar multiplication with the base point $G$. For resource-constrained IoT devices, computing two scalar multiplications continuously for each blockchain transaction is going to drain battery power of smart devices dramatically. Furthermore, containing an ephemeral public key in each stealth payment increases the size of the transaction and incurs additional communication overhead for IoT devices as well.

Motivated by the TLS session resumption techniques \cite{Die08,Sal08}, we aim to accelerate the process for receivers finding the matched destination address by extending the lifetime of the shared secret between senders and receivers. While both the session ID \cite{Die08} and the session ticket \cite{Sal08} are fixed in TLS for a given period of time between the client and server, the sender does need to generate a one-time destination address for each payment sent to the same recipient in our case. To this end, both the sender and receiver will apply the cryptographic hash function to their shared secret for subsequent $N$ transactions before the sender initiates a shared secret update with a fresh ephemeral public key. This key evolving process is shown in Fig. 2, which leads to the design of $\textsf{DKSAP-IoT}$, a faster dual-key stealth address protocol for blockchain-based IoT systems, as detailed in the next subsection.

\begin{figure}[ht]
\centering
\label{fig:keyevolve}
\includegraphics[width=\textwidth,height=3.7cm]{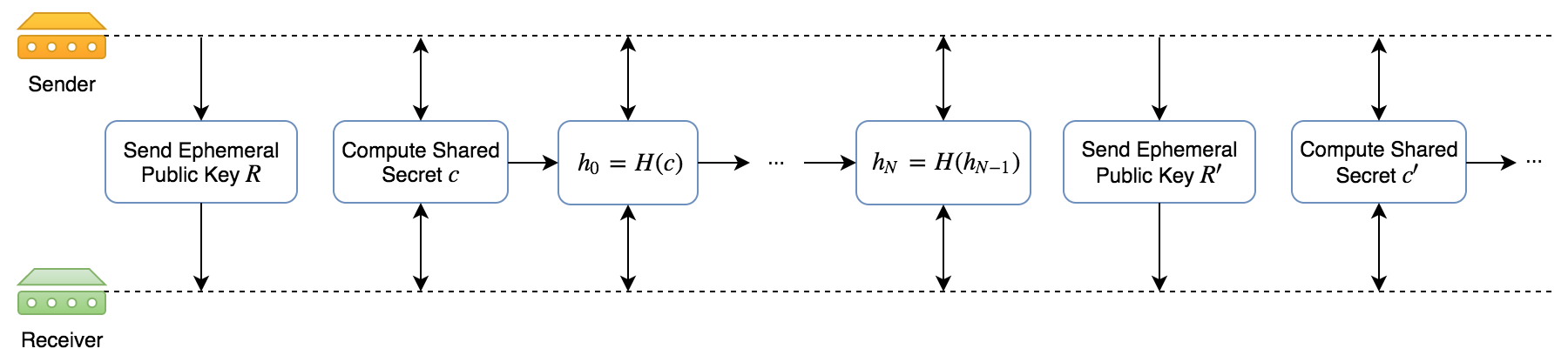}
\caption{The Key Evolving Process Between the Sender and Receiver in \textsf{DKSAP-IoT}}
\vspace{-0.7cm}
\end{figure}

\subsection{$\textsf{DKSAP-IoT}$ Specification}
$\textsf{DKSAP-IoT}$ is similar to $\textsf{DKSAP}$ except that whenever the sender and receiver establish a shared secret using ECDH it will be continuously and pseudorandomly updated with a cryptographic hash function and used in their subsequent $N$ stealth transactions. Both the sender and receiver maintain the transaction state (i.e., shared secret, counter, etc.) locally and update it after each stealth transaction. A high-level description of $\textsf{DKSAP-IoT}$ is depicted in Fig. 3.

\begin{figure}[ht]
\centering
\label{fig:stealthiot}
\includegraphics[width=\textwidth,height=9.5cm]{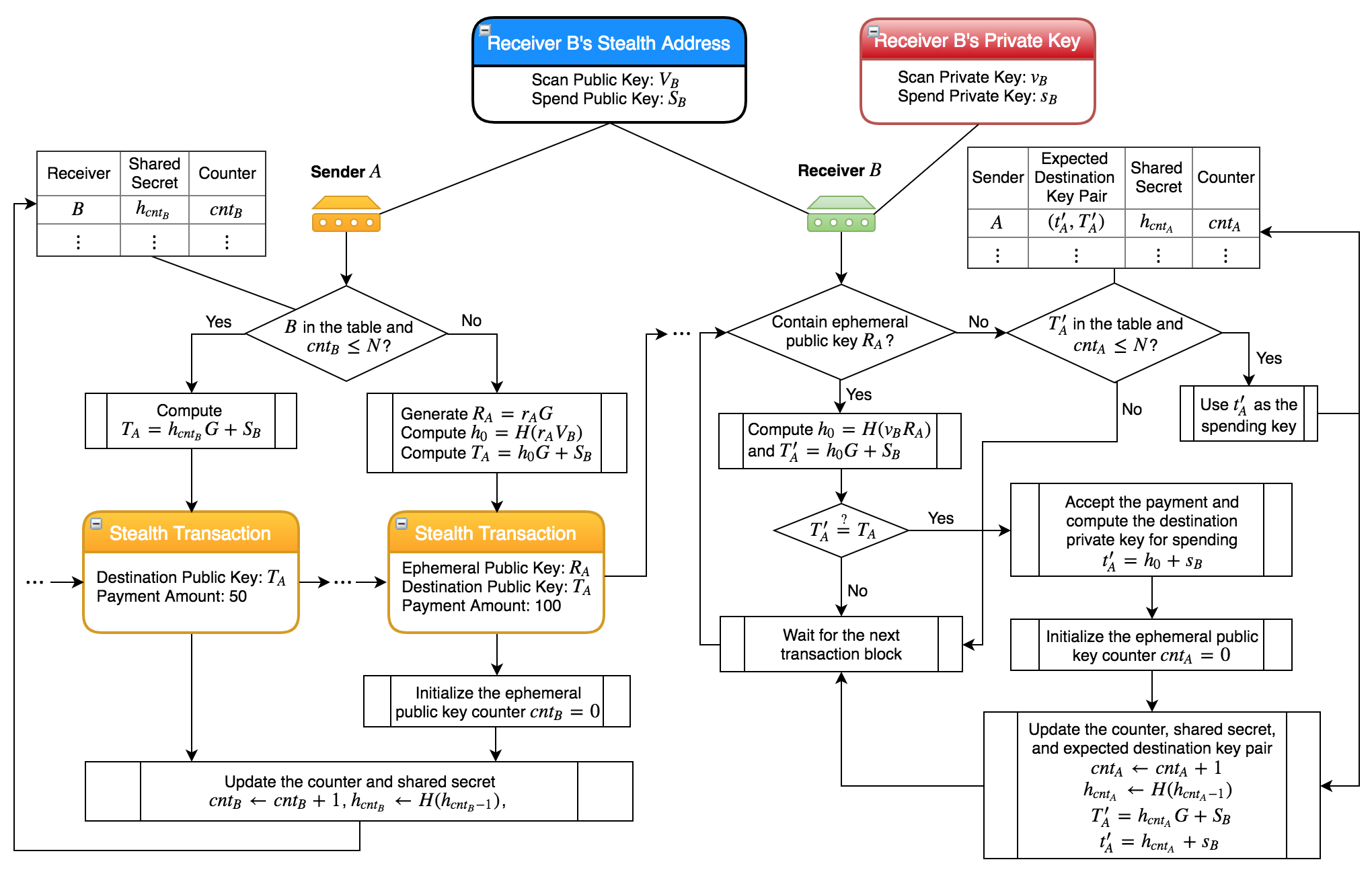}
\caption{The Dual-Key Stealth Address Protocol for IoT (\textsf{DKSAP-IoT})}
\end{figure}

In a blockchain-based IoT system, two smart devices $A$ and $B$ can process a transaction in a stealth mode using $\textsf{DKSAP-IoT}$ as described below:

\begin{enumerate}
\item The receiver device $B$ is pre-installed with a `scan key' pair $(v_B, V_B)$ and a `spend key' pair $(s_B, S_B)$ as in $\textsf{DKSAP}$, where $V_B = v_BG$ and $S_B = s_BG$.
\item For sending a transaction to $B$, the sender device $A$ first checks whether $B$ is in $A$'s receiver list. If $B$ is in the list and the counter value $cnt_B$ is less than $N$ (i.e., $A$ has communicated with $B$ before), $A$ retrieves the shared secret $h_{cnt_B}$ from the table and computes the destination public key: 
\begin{displaymath}
T_A = h_{cnt_B}G + S_B.
\end{displaymath}
The stealth transaction that only contains the destination public key $T_A$ as well as the payment amount is then added into the blockchain. In the case that $B$ is not in the list or the counter value is greater than $N$, $A$ generates a fresh ephemeral public key $R_A = r_AG$ and calculates the shared secret $h_0 = H(r_AV_B)$ as well as the destination public key as in $\textsf{DKSAP}$:
\begin{displaymath}
T_A = h_0G + S_B.
\end{displaymath}
Here the stealth transaction is composed of the ephemeral public key $R_A$, the destination public key $T_A$ and the payment amount. After putting the transaction on the blockchain, the sender $A$ will initialize the ephemeral public key counter $cnt_B = 0$. In both cases, the counter $cnt_B$ and the shared secret $h_{cnt_B}$ will be updated as well:
\begin{displaymath}
cnt_B \leftarrow cnt_B + 1,\;\;h_{cnt_B} \leftarrow H(h_{cnt_B - 1}).
\end{displaymath}
Note that only the counter $cnt_B$ is updated when it reaches $N$.
\item Upon receiving a stealth transaction, the receiver $B$ first checks whether the transaction contains an ephemeral public key $R_A$. If it is, $B$ computes the purported shared secret and destination public key:
\begin{displaymath}
h_0 = H(v_BR_A),\;\;T_A' = h_0G + S_B.
\end{displaymath}
If the purported destination public key $T_A'$ matches the received one (i.e., $T_A' = T_A$), $B$ accepts the payment from $A$ and computes the corresponding private key for spending:
\begin{displaymath}
t_A' = h_0 + s_B.
\end{displaymath}
$B$ also sets the ephemeral public key counter $cnt_A$ to be 0, updates the counter and shared secret, and precomputes the expected destination key pair for the next stealth transaction from $A$:
\begin{eqnarray}
cnt_A \leftarrow cnt_A + 1,\;\;h_{cnt_A} \leftarrow H(h_{cnt_A - 1}),\\
T_A' = h_{cnt_A}G + S_B,\;\;t_A' = h_{cnt_A} + s_B.
\end{eqnarray}
When $B$ receives a stealth transaction without an ephemeral public key, $B$ will check whether the received destination public key $T_A$ is contained in its list of senders. If a match is found and the value of the counter $cnt_A$ is less than or equal to $N$, $B$ retrieves the corresponding destination private key $t_A'$ as the spending key and updates the transaction state information accordingly with the equations (1) and (2). Again only the counter $cnt_A$ is updated when it reaches $N$.
\end{enumerate}

In \textsf{DKSAP-IoT}, stealth transactions are divided into two categories depending on whether ephemeral public keys are included in the blocks. For the first stealth transaction between two blockchain nodes, the receiver needs to conduct the same operations as \textsf{DKSAP}, followed by a more efficient preparation process for the next transaction. For the subsequent $N$ stealth transactions between the same peers, generating a fresh ephemeral key is no longer needed on the sender side. Meanwhile, the receiver only performs a fast table look-up as well as the transaction state updates, which facilitates the receiver to quickly filter out the designated transactions.

Given the `scan private key' $v_B$ and the `spend public key' $S_B$, the auditor/proxy server is able to calculate all the destination addresses for the receiver $B$, thereby tracking or forwarding all the transactions to $B$. However, both the auditor or the proxy server cannot derive the corresponding ephemeral private keys and spend the funds.   

\section{Security Analysis and Performance Evaluation}
\label{sec:performance}
 
In this section, we analyze the security and performance of \textsf{DKSAP-IoT} and report its implementation on a Raspberry Pi 3 Model B, a good representative of moderately resource-constrained embedded devices.  

\subsection{Security Analysis}
\textsf{DKSAP-IoT} follows the same threat model as \textsf{DKSAP}, in which the adversary aims to determine the corresponding recipients by observing the transactions on the blockchain. \textsf{DKSAP-IoT} provides the following security properties:
\begin{itemize}
\item \textbf{Receiver Anonymity}: \textsf{DKSAP-IoT} offers strong anonymity for receivers and ensures the unlinkability of payments received by the same payee. For each payment to a stealth address, the sender computes a new normal address $T_A$ on which the funds ought to be received. Given two destination addresses $T_A^{(i)} = h_iG + S_B$ and $T_A^{(j)} = h_jG + S_B$ ($0 \leq i, j \leq N$) for the same receiver $B$, the adversary is not able to link them thanks to the difficulty of ECDLP.      
\item \textbf{Forward Privacy}: \textsf{DKSAP-IoT} provides forward secrecy due to the usage of a cryptographic hash function for updating the shared secret continuously for $N$ stealth transactions. If the adversary compromises the device and obtains $h_l$ for the $l^{\textrm{th}}$ ($0 < l < N$) stealth transaction, he/she is still not able to link previous transactions because of the properties of the hash function. 
\item \textbf{Stealth Transaction Hiding}: In \textsf{DKSAP}, transactions in the stealth mode can be easily distinguished from regular ones in the blockchain due to the presence of ephemeral public keys, thereby resulting in some loss of privacy. However, the ephemeral public key only needs to be updated every $N$ stealth transactions for two communication peers in \textsf{DKSAP-IoT} and those stealth transactions in between are not distinguishable from regular ones.   
\end{itemize}

Since both the sender and receiver need to locally maintain the state information for their peers in \textsf{DKSAP-IoT} (See Fig. 3), these tables, together with the device private keys, should be stored in the encrypted form for mitigating the risk that IoT devices might get compromised. Considering that the hardware AES engine is widely available on many IoT devices, the computational overhead for encrypting/decrypting those sensitive information is quite small.        

\subsection{Performance Evaluation}
\subsubsection{Computational and Communication Overhead.} 
We assume that a sender is going to send $N$ stealth transactions to a receiver using blockchain. Let \textsf{RP}, \textsf{FP} and \textsf{H} denote the computation of a random-point scalar multiplication, a fixed-point scalar multiplication and a cryptographic hash function, respectively. Table 1 gives a comparison between the \textsf{DKSAP} and \textsf{DKSAP-IoT} in terms of their computational overhead.

\vspace{-0.5cm}
\begin{table}[h]
\centering
\caption{Computational Overhead of \textsf{DKSAP} and \textsf{DKSAP-IoT} for Sending $N$ Stealth Transactions between Two Blockchain Nodes}
\begin{tabular}{|c|c|c|c|c|c|c|}
\hline
\multirow{2}{*}{Scheme} & \multicolumn{3}{c|}{Sender} & \multicolumn{3}{c|}{Receiver} \\
\cline{2-7}
 & \#\textsf{RP} & \#\textsf{FP} & \#\textsf{H} & \#\textsf{RP} & \#\textsf{FP} & \#\textsf{H}\\
\hline
\textsf{DKSAP} & $N$ & $2N$ & $N$ & $N$ & $N$ & $N$ \\
\hline
\textsf{DKSAP-IoT} & $1$ & $N + 1$ & $N$ & $1$ & $N$ & $N$ \\
\hline
\end{tabular}
\end{table}

From Table 1, one can see that \textsf{DKSAP-IoT} is able to reduce the number of \textsf{RP} and \textsf{FP} by $N - 1$ on the sender side, respectively, when compared to the \textsf{DKSAP}. Moreover, \textsf{DKSAP-IoT} can also save $N - 1$ \textsf{RP} on the receiver side. With respect to the communication overhead, the sender in \textsf{DKSAP-IoT} only needs to contain a fresh ephemeral public key in the first stealth transaction, thereby saving the transmission of $N - 1$ elliptic curve points.

\subsubsection{Software Implementation.} To validate the performance improvements of \textsf{DKSAP-IoT}, we implemented an optimized elliptic curve cryptography library, namely \textsf{libsect283k1}\footnote{\textsf{libsect283k1} will be integrated into the IoTeX testnet and mainnet as part of the \textsf{iotex-core} (see \url{https://github.com/iotexproject/iotex-core}).}, using the $283$-bit binary Koblitz curve specified in~\cite{SEC2}:
\begin{displaymath}
E(\mathbb{F}_{2^{283}}): y^2 + xy = x^3 + 1,
\end{displaymath}
where the binary field $\mathbb{F}_{2^{283}}$ is defined by $f(x) = x^{283} + x^{12} + x^7 + x^5 + 1$. The library was written in C and compiled using the GNU C Compiler (GCC). A number of efficient techniques, such as the lambda coordinates~\cite{Oli:Lop:Ara}, the window $\tau$NAF method~\cite{Han03}, the pre-computation~\cite{Yu:Mus:Xu}, etc., have been utilized to optimized the performance of the \textsf{libsect283k1} library. Moreover, \textsf{BLAKE-256}~\cite{Saa:Aum} is chosen as the hash function in our library due to its high performance cross multiple computing platforms. When running our library on a Raspberry Pi 3 Model B, the timings for the computation of \textsf{RP}, \textsf{FP} and \textsf{H} are shown in Table 2. 

\vspace{-0.5cm}
\begin{table}
\centering
\caption{Timings for Computing \textsf{RP}, \textsf{FP} and \textsf{H} on a Raspberry Pi 3 Model B}
\begin{tabular}{|c|c|c|}
\hline
\textsf{RP} & \textsf{FP} & \textsf{H} \\
\hline
$3.67$ ms & $3.12$ ms & $5.26$ $\mu$s \\
\hline
\end{tabular}
\end{table}

Note that the computation of the hash function is about three orders of magnitude faster than that of the scalar multiplication over an elliptic curve. Therefore, using the hash function to update the shared secret and extend its lifetime brings significant performance benefits for IoT devices. Fig. 4 compares the performance of the \textsf{DKSAP} and \textsf{DKSAP-IoT} on both sender and receiver sides for sending $N = 10, 20$ and $30$ stealth transactions, respectively.    

\begin{figure}[ht]
\centering
\label{fig:performance}
\includegraphics[width=9cm, height=12cm]{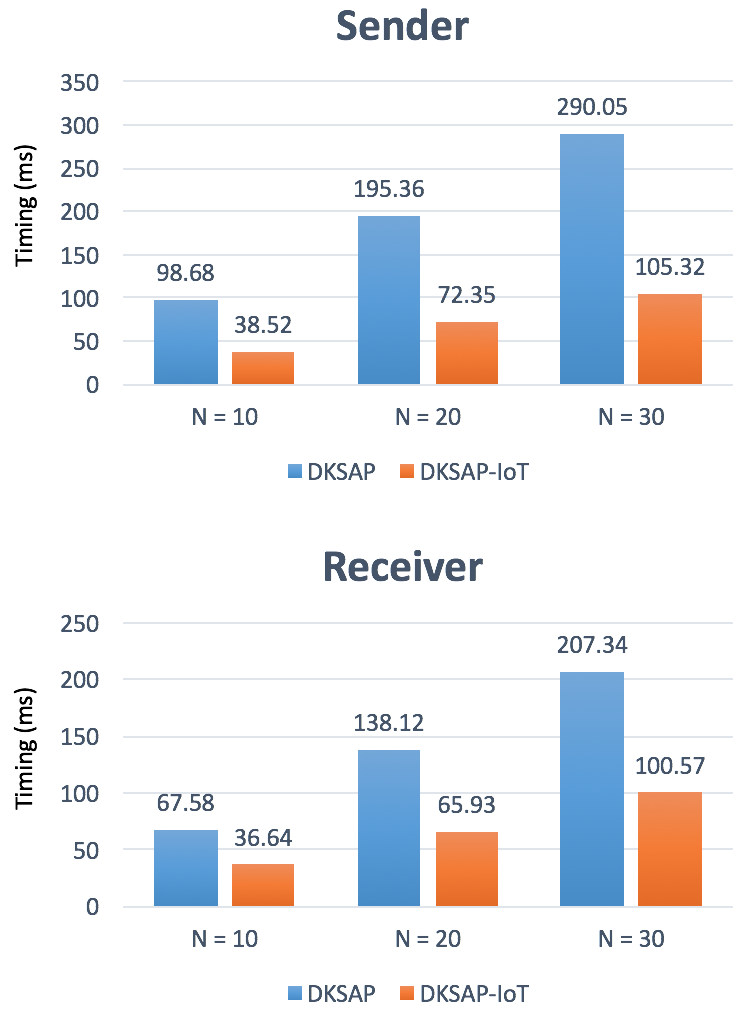}
\caption{Performance Comparison of the \textsf{DKSAP} and \textsf{DKSAP-IoT} for Sending $N = 10, 20$ and $30$ Stealth Transactions}
\end{figure}

From Fig. 4, one notices that the overall cost of \textsf{DKSAP-IoT} is less than 50\% of \textsf{DKSAP}, mainly because extending the lifetime of the shared secret with a cryptographic hash function enables both the sender and receiver to reduce the number of \textsf{RP} from $N$ to $1$. Moreover, the computation of the hash function is almost negligible compared to the scalar multiplication over the elliptic curve. In addition, \textsf{DKSAP-IoT} can save the transmission of $72\cdot(N - 1)$ bytes for $N$ stealth transactions. For resource-constrained IoT devices, the improved performance and reduced transaction size by \textsf{DKSAP-IoT} leads to significant power savings and extended battery life. 

\section{Conclusion}
\label{sec:conclusion}
In this paper, we propose an efficient dual-key stealth address protocol \textsf{DKSAP-IoT} for blockchain-based IoT systems. Motived by the TLS session resumption techniques, we apply a cryptographic hash function to continuously update a shared secret between two communication peers and extend the lifetime of this shared secret for additional $N$ transactions. Both the sender and receiver need to maintain the state information locally in order to keep track of the pairwise session keys. The security analysis shows that \textsf{DKSAP-IoT} provides receiver anonymity and forward privacy. When implementing \textsf{DKSAP-IoT} on a Raspberry Pi 3 Model B, we demonstrate that \textsf{DKSAP-IoT} can achieve at least 50\% performance improvement when compared to the original \textsf{DKSAP}, besides significant reduction of the transaction size in the block. Our work is another logic step towards providing strong privacy protection for blockchain-based IoT systems.

\section*{Acknowledgement}
The author would like to thank the IoTeX team for the great support during the course of writing this paper and the anonymous reviewers for their insightful comments. 

%
%

\end{document}